\documentclass[a4paper,11pt]{article}
 \usepackage[latin1]{inputenc}
\usepackage[T1]{fontenc}
\usepackage{amsmath}
\usepackage{amsfonts}
\usepackage{amssymb}
\usepackage{color}
\usepackage{graphicx}

   \newcommand{\be}{\begin{equation}}
	 \newcommand{\ee}{\end{equation}}
	 \newcommand{\ba}{\begin{eqnarray}}
		 \newcommand{\ea}{\end{eqnarray}}
		 
		   \newcommand{\bea}{\begin{eqnarray}}
			 \newcommand{\eea}{\end{eqnarray}}

\begin{document} \title{Gauged Lifshitz model with Chern-Simons term}
\author{Gustavo Lozano$^a$, Fidel Schaposnik$^b$\thanks{Also at CICBA}  and
Gianni Tallarita$^b$ \\ \vspace{0.2 cm} \\
{\normalsize \it  $^{a}$Departamento de F\'\i sica, FCEyN,
 Pabell\'on 1, Ciudad Universitaria}\\
{\normalsize \it  Universidad de Buenos Aires, 1428, Buenos Aires, Argentina}\\
~
\\
{\normalsize \it $^b$\it Departamento de F\'\i sica, Universidad Nacional de La Plata}\\ {\normalsize \it Instituto de F\'\i sica La Plata}\\ {\normalsize\it C.C. 67, 1900 La Plata, Argentina}}

\date{\hfill}

\maketitle
\begin{abstract}
We present a gauged Lifshitz Lagrangian including second and forth order spatial derivatives of the
scalar field and a Chern-Simons term, and study non-trivial solutions of the classical
equations of motion. While the coefficient $\beta$ of the forth order
 term should be positive in order to
guarantee positivity of the energy, the coefficient $\alpha$ of the quadratic one need not be.
We investigate the parameter domains finding significant differences in the field behaviors.
Apart from the usual vortex field behavior of the ordinary relativistic Chern-Simons-Higgs model, we
 find in certain parameter domains   oscillatory solutions reminiscent of
the modulated phases of Lifshitz systems.
\end{abstract}
\newpage
\section{Introduction}

 In their  well-honored proposal to describe dual strings \cite{NO},  Nielsen and Olesen  stressed  the connection between the Abelian Higgs model and the Ginzburg-Landau theory of superconductivity, relating the free energy in the latter with the action for static configurations of the former. In this way, the vortex filaments of type-II superconductors were identified with
 string-like classical solutions in a gauge theory with spontaneous symmetry breaking.

More than 30 years ago  Ginzburg proposed \cite{Ginzburg}
a  generalization of the Ginzburg-Landau functional by including higher derivative terms, this implying an anisotropic coordinate scaling, in order to describe superdiamagnets - a class of materials with strong
diamagnetism  but differing from conventional superconductors. Such
generalization was then used to analyze \cite{Buzdin} the properties of superconductors near a tricritical Lifshitz  point, a point in the phase diagram  at which a disordered phase, a spatially homogeneous
ordered phase and a spatially modulated ordered phase meet.

{
The study of Lifshitz critical points has recently attracted much attention, not only in connection with
condensed matter systems (see \cite{Diehl} and references therein) but also in the analysis of gravitational models in which anisotropic scaling leads to improved short-distance behavior (see \cite{Hori} and references therein). A link between these two issues was established in \cite{Kachru} within the framework of the gauge/gravity correspondence by searching gravity duals of nonrelativistic   quantum field theories with anisotropic scaling, dubbed in \cite{Fradkin} as ``Lifshitz field theories''.}

The question that we address in
this work is whether one can find Nielsen-Olesen like solutions when anisotropic scaling
 is introduced in the Abelian Higgs model through the addition of higher order spatial derivatives. As
 a laboratory we consider a $2+1$ dimensional model with a complex Higgs scalar coupled to a $U(1)$ gauge field with a Chern-Simons (CS) action  \cite{DJT}. The topological character of the CS term avoids the possibility  of including higher order derivatives for the gauge field action (as it would be case for the  Maxwell
action).

When  higher order derivative terms in the scalar Lagrangian are absent, the Chern-Simons-Higgs model has
vortex-like finite energy solutions carrying both   quantized magnetic flux $\Phi$ and non trivial electric charge
$Q =- \kappa \Phi$ with $\kappa$ the CS coefficient \cite{Paul}-\cite{deVega}. Moreover, for an appropriate sixth-order symmetry breaking potential,   first order BPS equations \cite{Cor}-\cite{Cugliandolo} exist, which can be easily found by analyzing the supersymmetric extension of the model \cite{Lee}. Our goal will be to determine whether this kind of solutions also exists in a ``Lifshitz Abelian Higgs'' model  and, in the affirmative, how they depend on the parameters associated to the Lagrangian scaling anisotropy.

The plan of the paper is the following: we introduce in section 2 a $(2+1)$-dimensional Lifshitz-Higgs model with gauge field dynamics governed by a Chern-Simons term. In order to solve the
 classical equations of motion we make the same ansatz leading to vortex solutions in the ordinbary (relativistic) case. Then, in section 3 we analyze the asymptotic behavior of the gauge and scalar fields resulting from the equations of motion, showing the existence of four regions according to  the
 values of the parameters of the model.
 We discuss in section 4 the properties of the solutions obtained numerically i giving a summary of results and a discussion on possible extensions of our work
  in section 5. We briefly
  describe in an appendix the linearized
 approximation we employed to determine the asymptotic behavior of the solutions in different
 parameter regions.

\section{The Lagrangian}
We consider a $2+1$ dimensional model with Chern-Simons-Higgs Lagrangian
\be
L= \gamma \left| D_0[A]\phi\right|^2
  - \alpha \left| D_i[A]\phi\right|^2   -\beta\left|D_i[A]D_i[A]\phi\right|^2  + V[|\phi|] +
\frac\kappa2 \varepsilon^{\mu\nu\alpha} A_\mu \partial_\nu A_\alpha
\label{Uno}\ee
with $\mu=0,1,2$ and $i=1,2$. The metric  signature is $(1,-1,-1)$.
We  consider space and time coordinate units so that
\be
[x]^2 = [t] \; .
\label{escala}
\ee
Accordingly, $\gamma$, $\beta$ and $\kappa$ are dimensionless and $\alpha$ has length dimensions
$[\alpha] =  {-2}$.  Concerning the dimensions of the complex scalar $\phi$ and $U(1)$ gauge field $A_\mu$ one has $[\phi] = 0$, $ [A_i] = {-1}$, $[A_0] = -2$.

The Lagrangian (\ref{Uno}) is a generalization of the one considered in \cite{Cor}-\cite{Jackiw:1990aw}
incorporating higher (forth) order covariant derivative terms for the scalar fields.
For vanishing potential and at the ``Lifshitz point'' $\alpha =0$, the Lagrangian  is invariant under anisotropic scaling
with ``dynamical critical exponent'' $z=2$
\be
x \to \lambda x \; , \;\;\; t \to \lambda^2 t \; .
\ee
Note that the choice of a Chern-Simons term ensures that scale invariance is preserved
even in the presence of gauge fields (as opposed to what would happen with a standard Maxwell term).

The covariant derivative $D_\mu$ acts on the scalar field $\phi$
according to
\be
D_\mu[A] \phi = (\partial_\mu + ieA_\mu)\phi
\ee
with $[e] = 0$. The potential $V[\phi]$  is to be  specified below.

Given the Lagrangian (\ref{Uno}) one gets  Gauss's law by differentiating with respect to $A_0$,
\be
\kappa \varepsilon^{0ij}\partial_iA_j = j^0
\label{gauss}
\ee
where
\be
j_0 = ie\gamma (\phi^* D_0\phi - \phi D_0\phi^* ) = -2e^2 {\gamma}A_0|\phi|^2
\label{current}
\ee
Defining
\be
B = -\varepsilon^{ij} \partial_iA_j
\label{b}
\ee
one then has, using eq.(\ref{gauss}),
\be
A_0 = \frac{\kappa}{2e^2 {\gamma}}\frac{B}{|\phi|^2}
\ee
Inserting this result in eq.(\ref{current}) one gets
\be
j_0 = -\kappa B
\ee
so that the usual Chern-Simons-Higgs model relation between charge $Q$ and magnetic flux $\Phi$ holds
\be
Q = \int d^2x j^0 = -\kappa \int d^2x B \equiv -\kappa \Phi
\ee
The energy density  $\bar{\cal E}$ associated to  Lagrangian (\ref{Uno})  is
\be
\bar{\cal E} =\alpha
  \left| D_i[A]\phi\right|^2 +     \beta\left|D_i[A]D_i[A]\phi\right|^2 + \frac{1}{4\gamma e^2} \frac{\kappa^2 B^2}{|\phi|^2} + V[|\phi|] \; .
\label{Dos}
\ee
A lower  bound for the energy requires $\beta$ to be positive while $\alpha$ can have any sign.

As stated before, in the $  \beta = 0, \gamma = 1$ relativistic case and for a sixth order symmetry breaking potential this theory is known to have, at the classical level, self-dual vortex solutions both in the Abelian case \cite{Cor}-\cite{Jackiw:1990aw} and in its non-Abelian extension \cite{Cugliandolo}.

In order to solve the Euler-Lagrange equations deriving from
Lagrangian (\ref{Uno})  we consider the static axially symmetric  ansatz
\ba
&& \phi = f(r)\exp(-in\varphi)\label{1}
\\
&&A_\varphi = - \frac{A(r)}{r}
\label{2}
\\
&& A_0 = A_0(r)
\label{3}
\ea
with $n \in \mathbb{Z}$.
 Given this ansatz   the magnetic and electric fields
read
\be
B(r) = \frac1r \frac{dA(r)}{dr} \; , \;\;\; E(r) = -\frac{dA_0}{dr}.
\ee
The
  equations  of motion take the form
\ba
&& -\frac\kappa{r} \frac{dA(r)}
{dr} + 2 \gamma e^2 A_0(r)f^2(r)= 0
\label{ala1n}\\
~\nonumber\\
&& \kappa    \frac{d A_0}{dr} +\frac{4e^2\beta}{r} \left(\frac{n}e + A\right)   f\left( \frac{d^2 } {dr^2} + \frac1r \frac{d }{dr} - \frac{e^2}{r^2} \left(\frac{n}e + A\right)^2  \right)f
- \alpha \frac{2e^2}{r} \left(\frac{n}e + A\right)   f^2 = 0\label{ala2n}\\
 ~ \nonumber\\
&& \beta\left( \frac{d^2 } {dr^2} + \frac{1}r \frac{d }{dr} - \frac{e^2}{r^2} \left(\frac{n}e + A\right)^2  \right) \left( \frac{d^2f } {dr^2} + \frac1r \frac{d f}{dr} - \frac{e^2}{r^2} \left(\frac{n}e + A\right)^2 f \right) \nonumber\\
&& - \alpha\left( \frac{d^2 f} {dr^2} + \frac1r \frac{d f}{dr} - \frac1{r^2} (n + eA)^2 f \right)- \gamma e^2A_0^2(r)  f =
  \frac{1}{2}\frac{\partial V}{\partial f}
  \label{ala3n}
\ea

The potential $V$ is in general chosen so as to allow for spontaneous symmetry breaking. In the  relativistic $2+1$ dimensional case the most general renormalizable self-interacting scalar potential
is sixth order and in fact to find first order BPS equations it should  be of this order and take the form \cite{Cor}-\cite{Jackiw:1990aw}
\be
V = \frac{e^4\tau}{8\kappa^2}f^2(f^2-v^2)^2
\label{V}
\ee
with $v$ the Higgs field vev  and the coupling constant $\tau$ has length dimensions $[\tau] = - 2$.
In the relativistic model first order self dual equations
exist at a certain value $\tau = \tau_{BPS}$ which would correspond in
 the present Lifshitz case to $\tau_{BPS}= 8 /\alpha^2 $.  From here on, and in order to compare the Lifshitz model results with those arising in the relativistic case, we shall take $V$ as given in (\ref{V}) and $\tau = \tau_{BPS}$.

\section{Asymptotic behavior}
 We start by discussing the conditions that we shall impose at the origin and
at the boundary. We choose as conditions at the origin those leading to regular solutions
 in the relativistic  case (see for example \cite{Cor}):
\ba
f(r) &=& f_0 r^{|n|} \nonumber\\
A_0(r) &=& a_0 + c_0 r ^{2|n|} \hspace{2 cm} r \to 0 \nonumber\\
A(r) &=& d_0 r^{2|n| + 2}
\label{esa}
\ea
Note that  a constant term $a_0$ in the $A_0(r)$ expansion is included
in order to achieve consistency of eq.(\ref{ala1n})  at the origin.
Coefficients
$a_0$ and $d_0$ are related  according to
\be
d_0 = \frac{e^2}{\kappa(|n|+1)}a_0f_0^2 \, .
\ee

Concerning  large $r$,  we write
\bea
f(r) &\approx&  v + h(r) \label{bea1}\\
  A(r) &\approx& -\frac{n}e + a(r) \;  \;\;\;  \;\;\; \;\;\;  \;\;\;  \;\;\; r \to \infty \label{bea2}\\
 A_0(r) &\approx& a_0(r)
\eea
with $h(r), a(r)$ and $a_0(r)$ small fluctuations. We then linearize the equations of motion which  reduce to
\be
  -\beta\; \nabla^2_r\nabla^2_r\; h(r)+\alpha\nabla^2_r\; h(r)-\sigma h(r)=0
\label{fi1}\ee
\be
 -\frac{1}r \frac{da(r)}{dr} + \gamma\mu a_0(r) = 0
\label{fi2}\ee
\be  \frac{da_0(r)}{dr} - \frac{\alpha \mu}{r} a(r)= 0
\label{fi3}
\ee
where
\be
\nabla^2_r = \frac{d^2}{dr^2}+\frac{1}{r}\frac{d}{dr},
\ee
\be
\sigma = \frac{e^4\tau v^4}{2\kappa^2} \; , \;\;\;\; \mu =
\frac{2e^2 v^2}{\kappa}
\ee
Eqs.(\ref{fi2})-(\ref{fi3}) can be written as two decoupled second order equations
\be
\frac{d^2a_0}{dr^2} + \frac1r \frac{da_0}{dr} - \alpha\gamma\mu^2a_0=0
\label{51}
\ee
\be
\frac{d^2a}{dr^2} -\frac1r \frac{da}{dr} - \alpha\gamma\mu^2 a = 0.
\label{americano}
\ee

First we deal with the scalar field behavior. After writing
\be
h(r)= \frac{h^0}{\sqrt{r}}\exp({q r}),
\ee
with $h^0$ a constant,
the solutions are determined from the equation
\be
q^2_{\pm}=\frac{1}{2\beta}\left(\alpha\pm\sqrt{\alpha^2-4\beta\sigma}\right).
\ee
The asymptotic behavior of the scalar field is then given by
\be
f(r) \approx v + \frac{h^0}{\sqrt{r}}\exp(-q_\pm r)
\label{tonto}
\ee

From the  results one can see that there is a critical value for $\beta$
\be
\beta_{\text{crit}}=\frac{\alpha^2}{4\sigma}
\label{44}
\ee
above which $q_\pm^2$ become imaginary.

In the region $\alpha> 0$ and $\beta<\beta_{\text{crit}}$ the solutions for $q_{\pm}$ are real.
In particular, for $\beta\sigma\ll \alpha^2$
\be
q_+^2 \approx \frac\alpha\beta \; , \;\;\;\; q_-^2 \approx \frac\sigma\alpha
\ee
Note that $q_-$ coincides with  the standard relativistic case solution where it plays the
role of the Higgs field mass \cite{Cor}.

Concerning the region  $\alpha\geq 0$ and $\beta>\beta_{\text{crit}}$ one has
\be
 q^2_{\pm}=\frac{1}{2\beta}\left(\alpha\pm 2i\sqrt{\beta\sigma}\sqrt{1-\frac{\alpha^2}{4\beta\sigma}}\right)
 \label{cuarenta}
\ee
which gives a complex solution. This region corresponds to underdamped oscillations of the Higgs field. We can write this as
\be
q^2_\pm=\frac{\sigma}{\beta} e^{\pm i\chi}
\ee
where
\be
\tan(\chi)=\frac{\sqrt{4\beta\sigma-\alpha^2}}{\alpha}.
\ee
The solution is therefore
\be
h=\frac{h^0\exp{(-\lambda r)}}{\sqrt{r}}\cos(k \hspace{1mm}r+\delta)
\ee
where
\be
\lambda =\sqrt{\frac{\sigma}{\beta}}|\cos(\chi/2)|\;, \quad k =\sqrt{\frac{\sigma}{\beta}}|\sin(\chi/2)|
\ee
where  $\delta$  is a constant phase.

We now consider  the case of $\alpha<0$. In this case for $\beta<\beta_{\text{crit}}$ we have that
\be
q^2_{\pm}=\frac{1}{2\beta}\left(-|\alpha|\pm\sqrt{\alpha^2-4\sigma\beta}\right)
\label{42}
\ee
which is always negative leading to oscillatory solutions with wavenumbers $|q_\pm|$.

Finally let us consider the $\beta>\beta_{\text{crit}}$ region where the solutions become
\be
q^2_\pm=\frac{\sigma}{\beta} e^{\mp i\chi}
\label{tontono}
\ee
leading for the scalar field behavior to a situation  similar to the case of $\alpha>0$ with $\beta>\beta_{\text{crit}}$.

Let us now study the asymptotic behavior of the gauge fields. For $\alpha >0$ the consistent asymptotic behavior is
 \bea
a_0(r) &\approx& \frac{a_{0\infty}}{\sqrt{r}} \exp(-\bar k  r)\nonumber\\
a(r) &\approx& a_\infty\sqrt{r} \exp(-\bar k  r)
\eea
Notice that in this region the asymptotic field behavior ensures finite energy  and
quantized  magnetic flux as in the relativistic case
\be
\Phi = \frac{2\pi}{e} n \; , \;\;\;  n\in \mathbb{Z}
\ee

In the $\alpha = 0$ case linearization leading to eqs.(\ref{51})-(\ref{americano})
is no longer valid. Instead, writing $a = \sqrt r g(r)$ and using  the gauge field equations of motion one gets a second order nonlinear equation for $g$
 compatible with bounded solutions at infinity. As will be discussed in next section, we do find
 a bounded numerical solution for $\alpha = 0$.

Concerning the  $\alpha <0$ region, one has
\bea
a_0(r) &\approx& \frac{a_{0\infty}}{\sqrt{r}} \sin(\bar k r +  \bar \varphi )\nonumber\\
a(r) &\approx& a_\infty \sqrt{r} \cos(\bar k r + \bar \varphi)
\label{notiene}
\eea
with
\be
\bar k = \sqrt{|\alpha|\gamma}\mu \; , \;\;\; a_{\infty} = - \sqrt{\frac{\gamma}{|\alpha|}} \,{a_{0\infty}}
\label{nonotiene}
\ee
The oscillatory behavior of configurations satisfying (\ref{notiene}) will require
the introduction of appropriate boundary conditions at a finite radius $R$.

\section{Solutions}

We shall present in this section numerical solutions of eqs.(\ref{ala3n}) satisfying the asymptotic condition discussed above. For definiteness we take  $n = 1$ and we shall fix  $\gamma = 1$ {(Since we are considering static solutions, changing gamma amounts to a redefinition of the scalar field coupling with $A_0$)}. In order to ensure positivity of the energy  we shall take $\beta > 0$. We shall separately consider    $\alpha \geq 0$  and $\alpha<0$ regions.  Following the discussion in the previous section, we shall distinguish regions with $\beta \lessgtr \beta_{crit}$.  The numerical procedure is based on a forth-order finite differences method  applied in the interval $(\epsilon,R)$ with $\epsilon$ close to the origin and $R$ large,  in combination with the behavior of fields close to the origin given by eq. (\ref{esa}).

\subsection{The $\alpha \geq 0$ region}
We  start by studying the  the $\alpha >0$, $\beta<\beta_{\text{crit}}$ region.  We give the results of our numerical calculation for $E$ and $B$ in
figures 1 and the scalar field in figure 2.

~

\centerline{\includegraphics[width=0.7\textwidth]{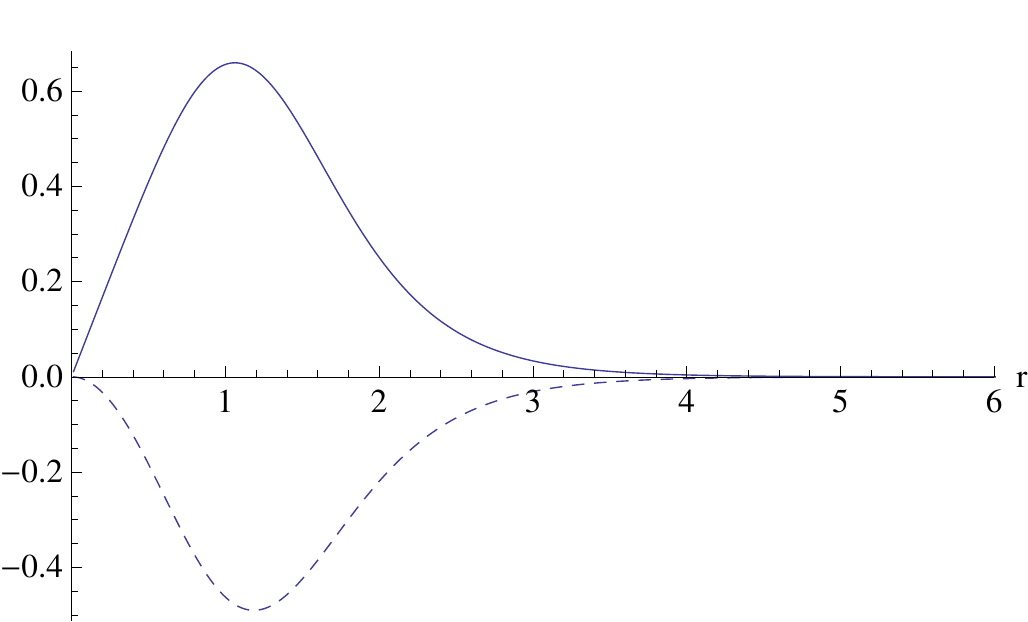}}
\noindent Figure 1: The electric (solid line) and magnetic (dashed line) fields in the region  $\alpha>0$, $\beta < \beta_{crit}$, with $\alpha = 1$, $\beta_{\text{crit}}= 0.0625$ and $\beta = 0.04$.  As in the relativistic Chern-Simons-Higgs model,   the magnetic and electric fields form a ring surrounding the vortex core.

~

One can see that the profile of the fields in this region exhibit slight deviations to
the relativistic case, originated
by the fourth order derivative terms. It should be noted that as $\beta$ grows we found numerically that the maximum magnitude of the
electric and magnetic fields decrease.
Concerning the Higgs field, it reaches its vacuum value exponentially according to eq.(\ref{tonto}), as can be seen in figure 2 with a similar profile as that corresponding to the relativistic case as shown in figure 2.

~

\centerline{\includegraphics[width=0.7\textwidth]{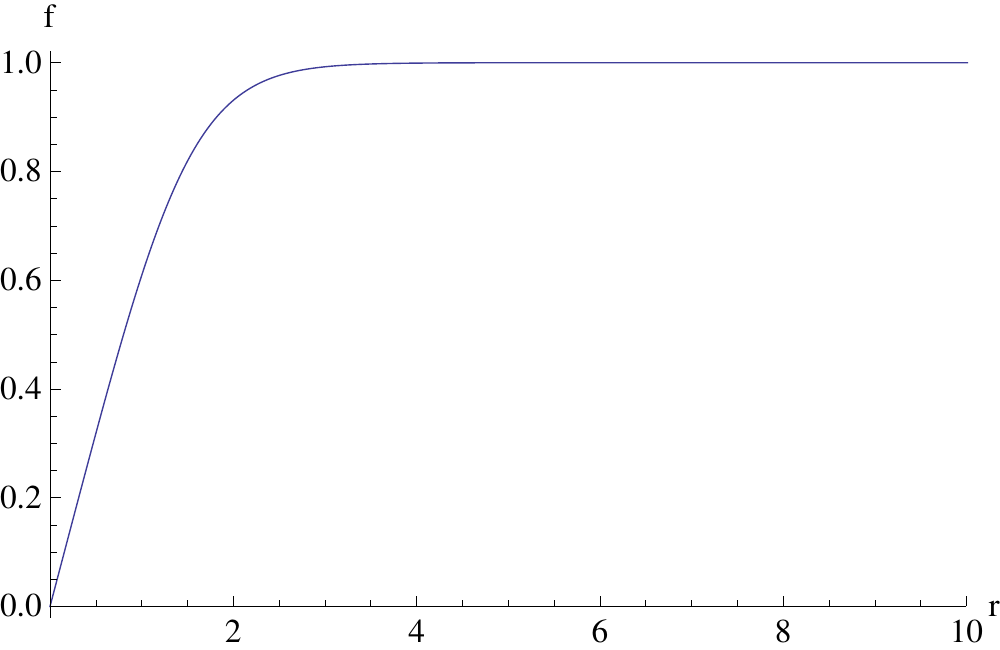}}
Figure 2: The Higgs field profile in the region corresponding to $\alpha>0$, $\beta < \beta_{\text{crit}}$. ($\alpha = 1, \beta_{\rm crit} = 0.0625$ and $\beta = 0.04$)

~

~

Let us  now  consider  $\beta>\beta_{\text{crit}}$ range where
  the  roots $q_{\pm}$ are complex (\ref{cuarenta}), this giving rise to underdamped oscillations in the Higgs the profile,
as shown in   figure (3).

~

\centerline{\includegraphics[width=0.7\textwidth]{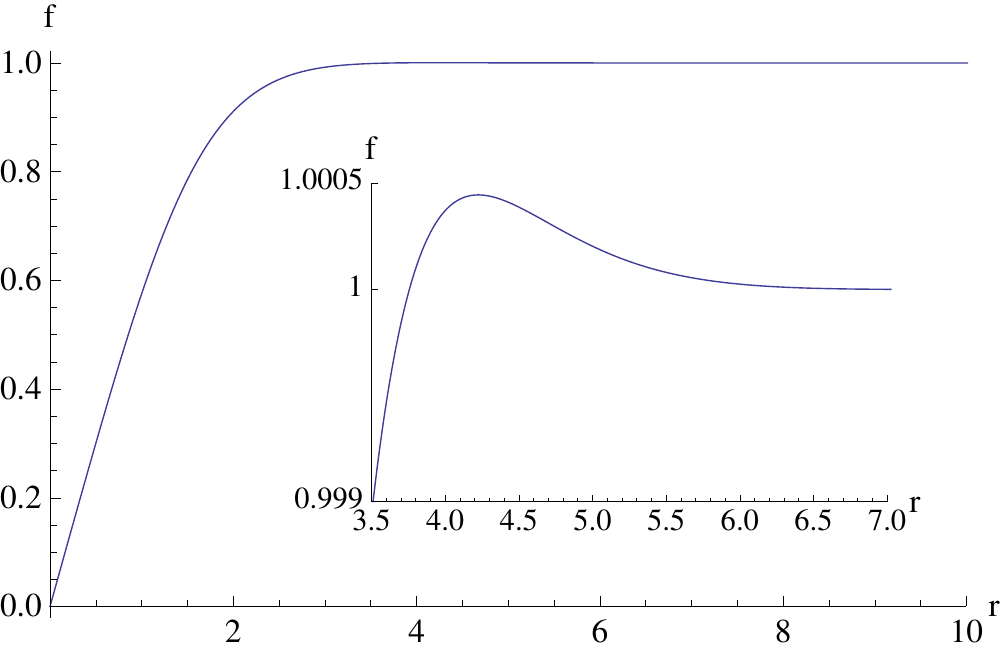}}
\noindent {Figure 3:  The Higgs field profile in the region $\alpha >0$, $\beta > \beta_{\rm crit}$. We have
chosen $\alpha = 1$, $\beta_{\rm crit} = 0.0625$ and $\beta = 0.2$. The inset shows a zoom  of the
 region where $f$ overshoots its vev and comes back to it, as is characteristic of an underdamped behavior}.

 ~

 For a given value of $\alpha$ the magnetic and electric field solutions  for $\beta >
 \beta_{\text crit}$ are qualitatively the same as those shown in figure 1 for $\beta <
 \beta_{\text crit}$.

 We then conclude that in the $\alpha >0 $ region the electric and magnetic field behavior is very similar to the ordinary relativistic CS-Higgs model. Concerning
 the scalar field, as one crosses from $\beta < \beta_{\rm crit}$ to $\beta > \beta_{\rm crit}$, it changes from the usual to an underdamped approach to its
 vacuum expectation value.

 We have studied the $\beta$-dependence of the energy in this region finding a linear behavior for small $\beta$. As an example, we show in figure 4 a numerical calculation of the energy ${\cal E}$ as a function of $\beta$ for
$\alpha = 1, \beta_{\rm crit} = 0.0625$. We find that ${\cal E}$ behaves approximately as  $
{\cal E} \approx {\cal E}_0 + 0.25\beta$.

 ~

\centerline{\includegraphics[width=0.7\textwidth]{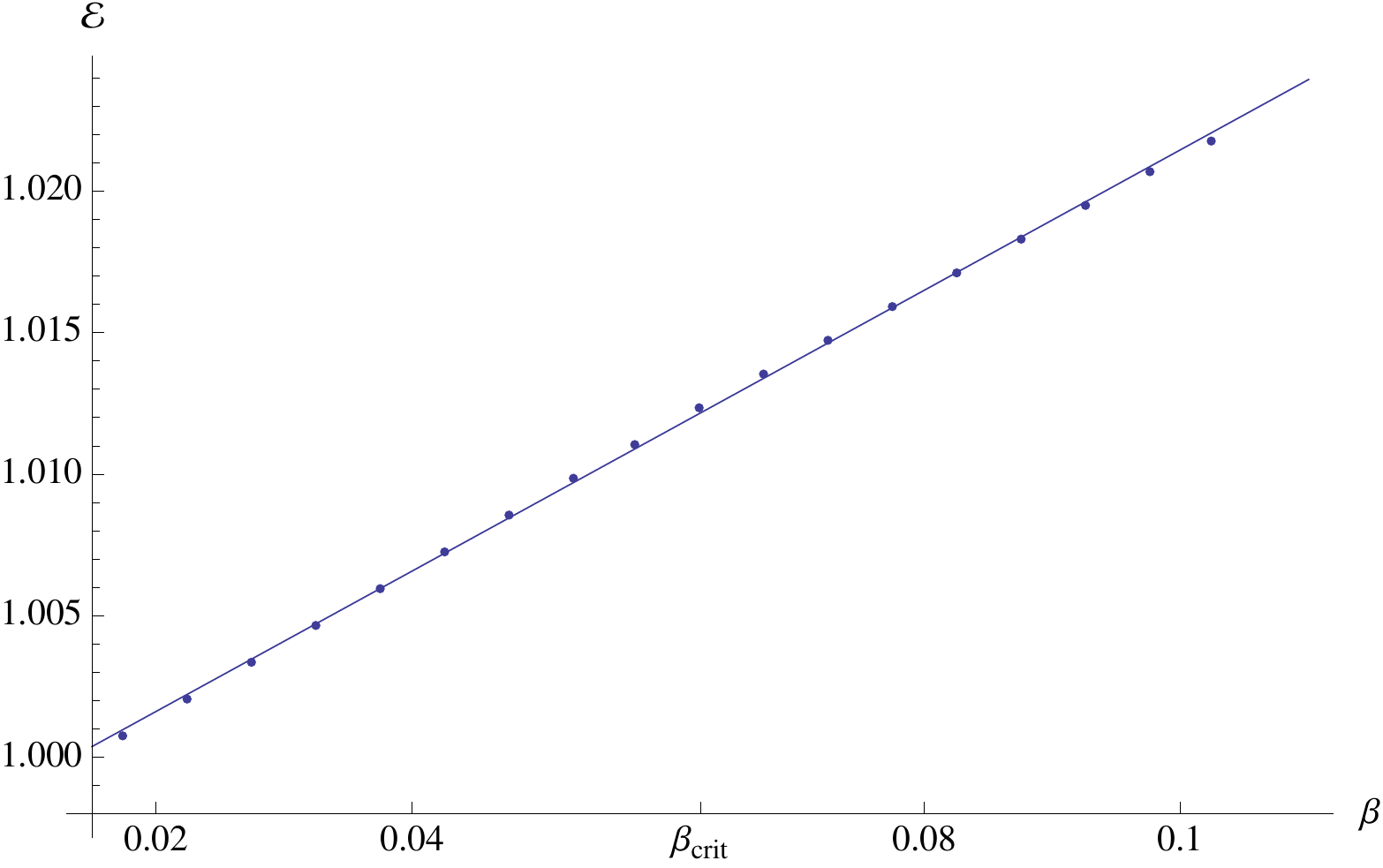}}
\noindent {Figure 4: The energy as a function of $\beta$ for $\alpha = 1,  \beta_{\rm crit} = 0.0625$}

~

We end this subsection by discussing   the  $\alpha = 0$ case for  which, for vanishing potential,
 the Lagrangian is invariant under anisotropic scaling with ``dynamical critical
exponent'' $z = 2$. In this case $\beta_{\rm crit}  = 0$ so that for any
$\beta>0$ the Higgs field shows an underdamped behavior.
We have numerically confirmed this result and also found bounded solutions for the gauge fields. The field profiles are qualitatively  similar to those found for $\alpha >0, \beta > \beta_{crit}$.

\subsection*{The $\alpha<0$ region}
One expects in this region a clearly different behavior compared to the relativistic CS-Higgs system since the negative sign of $\alpha$ in the $|D_i \phi|^2$ energy term implies not only
a change of sign in the $|\nabla \phi|^2$ term but also in the gauge field ``mass term'' that now has the ``wrong'' sign.

We start by studying  the $\beta>\beta_{\text{crit}}$ region where the fields asymptotic behavior is given by eqs. (\ref{tontono})-(\ref{nonotiene}). This behavior
leads to an  oscillatory energy density (and consequently to an in general unbounded energy).
For
example,  the third term in expression (\ref{Dos}) for the energy density takes the asymptotic form
\be
{\cal E}_3 = \frac{1}{4\gamma e^2} \frac{\kappa^2 B^2}{|\phi|^2}   \approx |\alpha| a_{0\infty}^2 e^2v^2 \frac{\sin^2(\bar kr + \bar \varphi)}{r}
\ee

{We show in  figure 5 the electric and magnetic fields in the $\alpha <0, \beta > \beta_{\rm crit}$ region.  Their  profiles show  the asymptotic oscillatory damped behavior consistent with eq.(\ref{notiene}). The behavior of the scalar field is
presented in figure 5. A zoom   outside the vortex core shows  damped oscillations consistent with equations (\ref{tonto})-(\ref{tontono}).}

 ~

 \centerline{\includegraphics[width=0.7\textwidth]{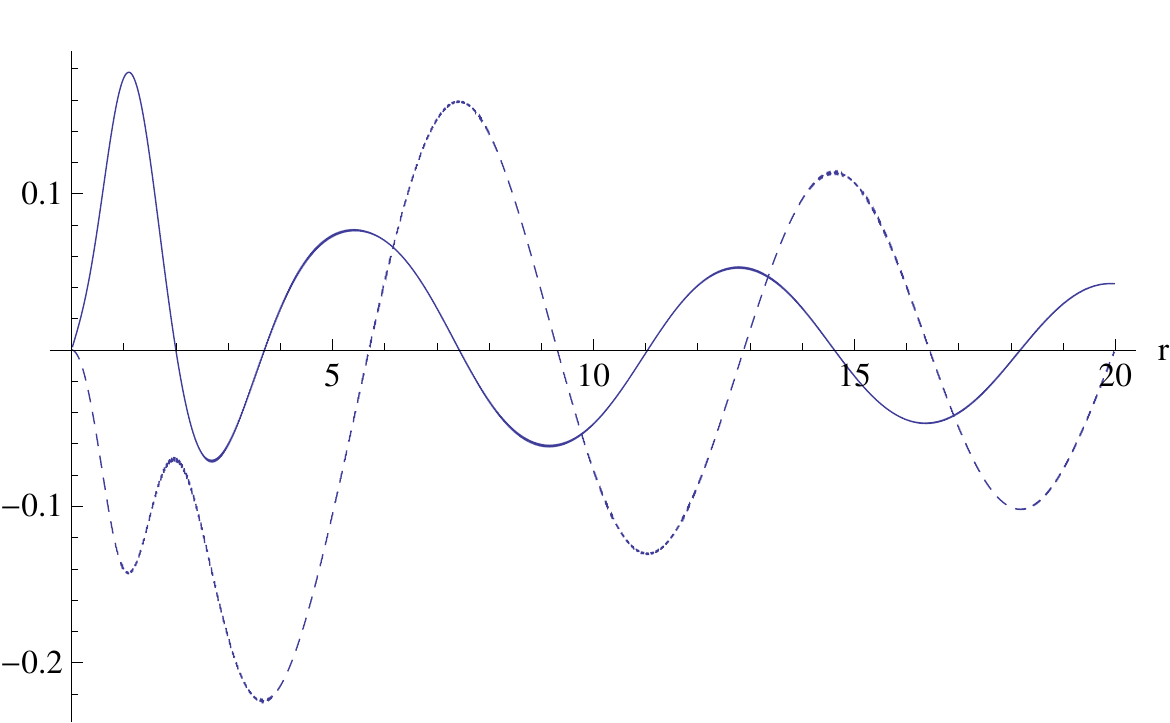}}
\noindent Figure 5: The electric (solid line) and magnetic (dashed line) fields in the region  $\alpha<0$, $\beta > \beta_{crit}$, with $\alpha = -0.2$, $\beta_{\text{crit}}= 0.0025$ and $\beta = 0.25$.

 ~

 ~

\centerline{\includegraphics[width=0.7\textwidth]{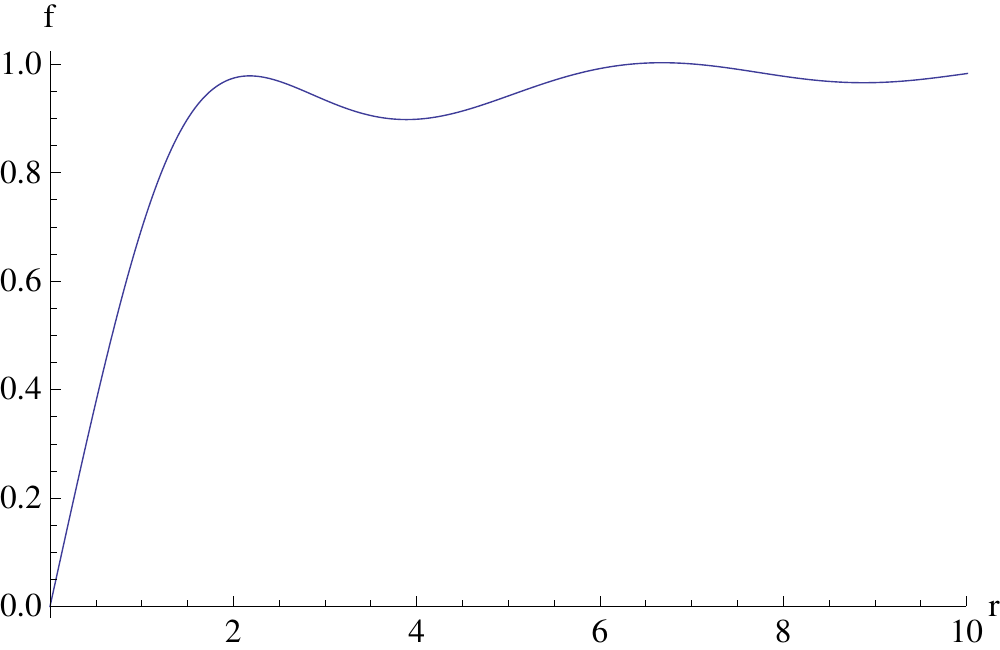}}
\noindent Figure 6: The Higgs field profile in the region  $\alpha<0$, $\beta > \beta_{crit}$, with $\alpha = -0.2$, $\beta_{\text{crit}}= 0.0025$ and $\beta = 0.25$.

 ~

 ~

In the  $\beta<\beta_{\text{crit}}$  region, the roots we found  in section 3, eq.(\ref{42}), lead to pure oscillatory solutions with no damping. The assumption of $h$ in eq.(\ref{bea1}) being asymptotically a
small perturbation to the scalar vacuum expectation value $v$ is then not self-consistent.  We have not been able to find stable solutions
of our $2+1$ model with the ansatz (\ref{1})-(\ref{2}). We indeed know that   in the absence of dynamical gauge fields this range of parameters
corresponds to the modulated ordered Lifshitz phase associated to spontaneous breaking of translations \cite{Michelson}. We then conclude that in this region a more detailed numerical study allowing the implementation of more general ans\"atze would be necessary.

\section{Summary and Discussion}
We have proposed  a gauged Lifshitz  Lagrangian with higher (forth) order spatial derivatives of the scalar field and a CS term and studied numerically non-trivial solutions of the classical equations of motion.
Notice that contrary to previous analysis of Lifshitz theories with CS term \cite{Mull} with  $z = 2$
we considered  higher derivatives for the scalar field rather than for the
gauge fields. As a consequence, the classical solutions of our model have a different character of the
ones resulting from such model \cite{SLGS}.

Coming back to the model we analyzed, let us recall that $\beta$, the coefficient of the forth order derivatives term,  was taken positive in order to ensure positivity of the energy. In contrast, the $\alpha$ coefficient
 multiplying the ordinary second order derivative term  could take both positive and negative values  being $\alpha = 0$ the Lifshitz point at which
 the model exhibits $z=2$
 anisotropic scaling  in the absence of a potential term.

 In order to solve the equations of motion we have made the static axially symmetric ansatz  that leads  to vortex solutions in the relativistic case.
For $\alpha > 0$ we have found solutions with magnetic and electric fields qualitatively similar to those of the ordinary relativistic model. The magnetic flux is quantized and the usual relation between electric charge and magnetic field in CS systems holds.
The difference with  the standard relativistic case manifests more  pronouncedly
 in the Higgs field behavior which for $\beta>\beta_{crit}$  approaches  its vacuum expectation value with  underdamped oscillations.
 The critical value is given by formula (\ref{44}), $\beta_{crit} = \alpha^2/4\sigma^2$, showing a dependence on the coefficient of the quadratic derivative coefficient and
on the parameters of the model (the value $v$ of the Higgs field at the minimum, the gauge coupling $e$, the CS coefficient $\kappa$ and the Higgs field self-interaction coupling constant $\tau$). For $\alpha = 0$ the numerical solutions that we found are qualitatively similar to those found for $\alpha >0,
\beta > \beta_{\rm crit}$.

The situation for the $\alpha <0$ region radically changes basically  because of the change in sign of the gauge field  mass term.  The ansatz led  to pure oscillatory solutions for the gauge fields with no damping.  Concerning the scalar field  one can again distinguish   two  situations depending on wether  $\beta$ is larger or smaller than
$\beta_{\rm crit}$.  In the former case we were able to find solutions exhibiting  electric and magnetic field
profiles with an asymptotic oscillatory  behavior while  the Higgs field profile shows damped oscillations. This behavior leads  in general to an oscillatory energy density and  an unbounded energy.  In the $\beta < \beta_{crit}$  region, the proposed ansatz led  to pure oscillatory solutions with no damping.

  We think that in the region $\alpha <0$ other terms in the Lagrangians,  as those considered by Ginzburg for  the  free energy of superdiamagnets and superconductors \cite{Ginzburg} might become relevant.  Also, more general ans\"atze, not purely relying  in cylindrical symmetry should be considered   in order
to incorporate the possibility of asymptotic   breaking of translational symmetry which is characteristic of modulated  Lifshitz phases. We hope to come back to this problem in a future work.

\section*{Appendix: The Fr\"obenius Method}
In this section we wish to apply Fr\"obenius's method to the linearized Higgs field equation of motion in order to determine its behaviour close to $r=0$, where $f$ is assumed to be small (see eq.(\ref{esa})) and the equation has a regular singular point. Following ref.\ \cite{Frobenius} we recast the equation of motion  (\ref{ala2n}) for the  Higgs field close to $r=0$ in simplified form as
\be
-\beta f'''' -\frac{2\beta}{r}f''' +\frac{(3\beta+\alpha r^2)}{r^2}f'' + \frac{(-3\beta+\alpha r^2)}{r^3}f' +
\left(\frac{\sigma}{4} v^4+\gamma e^2a_0^2 \right)f+\frac{(3\beta-\alpha r^2)}{r^4}f
=0
\label{eomf}
\ee
where we take the vorticity $n=1$ and ignore the contribution from $A(r)$ given that this vanishes at the origin. Note   that  higher order terms in $f$ coming from the potential are to  be ignored in the
linearized analysis. We proceed to make a Fr\"obenius ansatz for the behaviour close to the origin of the form
\be\label{ansatz}
f_{(\lambda)}=\sum_{m=0}^\infty {\cal F}_m (\lambda)r^{m+\lambda}.
\ee
Upon substituting this ansatz in eq.(\ref{eomf}) and looking at the lowest order in $r$ one obtains the indicial equation of the system, hence we look at the equation at order $r^{\lambda-4}$ where we obtain
\be
(\lambda-3)(\lambda-1)^2(\lambda+1)=0.
\ee
Therefore we have three distinct roots $\lambda=3,1,-1$ with multiplicities $1,2,1$ respectively. We proceed to determine the coefficients $a_m$ by looking at higher orders in $r$. The equation at order $r^{\lambda-3}$  implies that ${\cal F}_1=0$. The order $r^{\lambda-2}$ equation leads to
\be
\label{recursion}
{\cal F}_{2}(\lambda)=\frac{-\alpha {\cal F}_0}{\beta(1+\lambda)(3+\lambda)}
\ee
which gives solutions for both roots $\lambda=1$ and $\lambda=3$ as
\be
f_{(1)} = r\sum_{m=0}^\infty {\cal F}_m(1)r^{m}, \quad     f_{(3)}=r^3\sum_{m=0}^\infty {\cal F}_{m}(3) r^{m}
\ee
where $a_2$ and $b_2$ are coefficients extracted from eq.(\ref{recursion}) with the appropriate choice for $\lambda$, and an ill-defined solution for $\lambda=-1$ which we will return to later. The solution $f_2$ corresponds to the behaviour used in eq.(\ref{esa}) at $n=1$. Both these solutions and their derivatives are well behaved at the origin. The next order coefficients can be extracted from the order $r^\lambda$ equation as
\be
{\cal F}_4 (\lambda) = -\frac{{\cal F}_0\left(\alpha^2(1+3\lambda+\lambda^2)-\gamma e^2a_0^2\beta(3+4\lambda+\lambda^2)\right)}{\beta^2(1+\lambda)(3+\lambda)^2(7+17\lambda+8\lambda^2+\lambda^3)}
\ee
with higher order ${\cal F}_m$'s for odd $m$ vanishing. Being $\lambda=1$   a multiplicity 2 root, we know that the $\lambda$ derivative  of this solution is also a solution of the
equations of motion. In general if $f_{(\lambda)}$ is a solution of the form eq.(\ref{ansatz}), then
\be
\frac{df_{(\lambda)}}{d\lambda}=\ln{r} f_{(\lambda)}+r^\lambda\sum_{m=0}^\infty \frac{d{\cal F}_m(\lambda)}{d\lambda} r^{m}
\ee
which means that an independent solution is of the form
\be
\bar f_{(1)}=\frac{df_{(1)}}{d\lambda} =r\sum_{m=0}^\infty {\cal F}_m(1) r^{m}\ln{r}+r\sum_{m=0}^\infty \frac{d{\cal F}_m(1)}{d\lambda} r^{m}.
\ee
This solution has a singular derivative   at the origin.

The  solution of the linearized problem for $\lambda = -1$  takes the form
\be
f_{(-1)} = \frac{1}{r}\sum_{m=0}^\infty {\cal B}_{m} r^{m}+r\sum_{m=0}^\infty {\cal C}_{m} r^{m}\ln{r}
\ee
where as before the sum extends over even $m$  and  one finds that ${\cal B}_0$ and ${\cal C}_0$ are non-vanishing. This solution of the linearized problem diverges at $r=0$ and hence should not be taken into account
for searching physically acceptable solutions.

~

\noindent\underline{Acknowledgments}:   This work was supported by  CONICET  , ANPCYT , CIC, UBA and UNLP, Argentina.


\begin{thebibliography}{99}
\bibitem{NO}
  H.~B.~Nielsen and P.~Olesen,
  Nucl.\ Phys.\ B {\bf 61} (1973) 45.
  \bibitem{Ginzburg} V.~L.~Ginzburg, Pis'ma Zh\ Eksp.\ Fiz. {\bf 30} (1979) 345 (JETP Letters {\bf 30} (1979) 319).
      \bibitem{Buzdin} A.~I.~Buzdin and M.~L.~ Kuli\'c, J.\ of Low Temp.\ Phys.\ {\bf 54} (1983) 203.
\bibitem{Diehl} H.~W.~Diehl, Acta Phys. Slov. {\bf 52} (2002) 271.
\bibitem{Hori}
  P.~Horava,
  Class.\ Quant.\ Grav.\  {\bf 28} (2011) 114012
  [arXiv:1101.1081 [hep-th]].
  \bibitem{Kachru}
  S.~Kachru, X.~Liu and M.~Mulligan,
  Phys.\ Rev.\ D {\bf 78} (2008) 106005
  [arXiv:0808.1725 [hep-th]].
\bibitem{Hornreich} R.~M.~Honreich, M.~Luban and S.~Shtrikman, Phys.\ Rev.\ Lett.\
{\bf 35 } (1975)  1678.
   \bibitem{Fradkin} E.~Ardonne, P.~Fendley  and E.~Fradkin,
   Ann.\ of Phys.\ (N.Y.) 310 (2004) 493
  Phys.\ Rev.\ Lett.\  {\bf 66} (1991) 276.
  \bibitem{DJT}
  S.~Deser, R.~Jackiw, S.~Templeton,
  Phys.\ Rev.\ Lett.\  {\bf 48 } (1982)  975;
  Annals Phys.\  {\bf 140 } (1982)  372-411.
  \bibitem{Paul}
  S.~K.~Paul and A.~Khare,
  Phys.\ Lett.\ B {\bf 174} (1986) 420
   [Erratum-ibid.\  {\bf 177B} (1986) 453].
   \bibitem{deVega}
  H.~J.~de Vega and F.~A.~Schaposnik,
  Phys.\ Rev.\ Lett.\  {\bf 56} (1986) 2564;
  Phys.\ Rev.\ D {\bf 34} (1986) 3206.
  \bibitem{Cor}
  J.~Hong, Y.~Kim and P.~Y.~Pac,
  Phys.\ Rev.\ Lett.\  {\bf 64} (1990) 2230.
  \bibitem{Jackiw:1990aw}
  R.~Jackiw and E.~J.~Weinberg,
  Phys.\ Rev.\ Lett.\  {\bf 64} (1990) 2234.
 \bibitem{Cugliandolo}  L.~F.~Cugliandolo, G.~Lozano, M.~V.~Manias and F.~A.~Schaposnik,
  Mod.\ Phys.\ Lett.\ A {\bf 6} (1991) 479.
  \bibitem{Lee}
  C.~-k.~Lee, K.~-M.~Lee and E.~J.~Weinberg,
  Phys.\ Lett.\ B {\bf 243} (1990) 105.
  \bibitem{Gri} G.~Grinstein, Phys.\ Rev.\  B {\bf 23}, 4615 (1981) .
  \bibitem{Horava0}
  P.~Ho\v{r}ava,
  Phys.\ Lett.\ B {\bf 694},172 (2010)
  [arXiv:0811.2217 [hep-th]].
  \bibitem{Mulligan}
  M.~Mulligan, C.~Nayak and S.~Kachru,
  Phys.\ Rev.\ B {\bf 82} (2010) 085102
  [arXiv:1004.3570 [cond-mat.str-el]].
  \bibitem{Michelson} A.~Michelson, Phys.\ Rev.\ B {\bf 16}, 577 and 585 (1977).
  \bibitem{Frobenius} E.~A.~ Coddington and N. Levinson, {\it Theory of Ordinary Differential Equations}, Tata McGraw-Hill Pub. Co., New Dehli, 1972.
      \bibitem{Mull}  M.~Mulligan, C.~Nayak and S.~Kachru,
  Phys.\ Rev.\ B {\bf 82}, 085102 (2010)
  [arXiv:1004.3570 [cond-mat.str-el]].
  \bibitem{SLGS} I.~S.~Landea, N.~Grandi and G.~A.~Silva,
  arXiv:1206.0611 [cond-mat.str-el].
\end{thebibliography}
\end{document}